\documentclass[11pt,letter]{article}
\usepackage[hidelinks]{hyperref} 
\usepackage{fullpage}
\usepackage[utf8x]{inputenc}
\usepackage{amsthm}
\usepackage{todonotes,wrapfig,float,graphicx,amssymb,textcomp,array,amsmath}
\usepackage{enumerate,enumitem}
\usepackage{multirow}
\usepackage{tabularx}
\usepackage{color,xcolor}
\usepackage{enumitem}
\usepackage{float}
\usepackage{nicefrac}
\usepackage{version} % for \begin{comment}

\definecolor{mycolor}{rgb}{0, 0, 0}
\usepackage[titletoc,title]{appendix}

\usepackage{xparse}
\NewDocumentCommand\set{sm}{\IfBooleanTF#1{\{{#2}\}}{\left\{{#2}\right\}}}
\NewDocumentCommand\ceil{sm}{\IfBooleanTF#1{\lceil{#2}\rceil}{\left\lceil{#2}\right\rceil}}
\NewDocumentCommand\floor{sm}{\IfBooleanTF#1{\lfloor{#2}\rfloor}{\left\lfloor{#2}\right\rfloor}}
\NewDocumentCommand\round{sm}{\IfBooleanTF#1{\lfloor{#2}\rceil}{\left\lfloor{#2}\right\rceil}}
\NewDocumentCommand\pare{sm}{\IfBooleanTF#1{({#2})}{\left({#2}\right)}}
\NewDocumentCommand\range{smm}{\IfBooleanTF#1{\set*{{#2},\dots,{#3}}}{\set{{#2},\dots,{#3}}}}

\usepackage{lineno}
\makeatletter
\let\@fnsymbol\@arabic
\makeatother

\title{An Improved Bound for Plane Covering Paths}

\author{Hugo A. Akitaya\thanks{Miner School of Computer \& Information Sciences, University of Massachusetts, Lowell, USA, \texttt{hugo\_akitaya@uml.edu}. Supported by the NSF award CCF-2348067.} 
\and Greg Aloupis\thanks{Khoury College of Computer Sciences, Northeastern University, Boston, MA, USA, \texttt{g.aloupis@northeastern.edu}.} 
\and Ahmad Biniaz\thanks{School of Computer Science, University of Windsor, Windsor, Canada, \texttt{abiniaz@uwindsor.ca}. Supported in part by NSERC.}
\and Prosenjit Bose\thanks{School of Computer Science, Carleton University, Ottawa, ON, Canada, \texttt{jit@scs.carleton.ca, michiel@scs.carleton.ca}. Supported in part by NSERC.} 
\and Jean-Lou De Carufel\thanks{School of Electrical Engineering and Computer Science, University of Ottawa, Ottawa, ON, Canada, \texttt{jdecaruf@uottawa.ca,
leotheocharous3@gmail.com}. Supported in part by NSERC.} 
\and Cyril Gavoille\thanks{LaBRI, University of Bordeaux, France, \texttt{gavoille@labri.fr}. Supported by the French ANR projects ANR-22-CE48-0001 (TEMPOGRAL) and ANR-24-CE48-7768-01 (ENEDISC).}
\and John Iacono\thanks{Department of Computer Science, Universit{\'e} libre de Bruxelles, Brussels, Belgium. This work was supported by the Fonds de la Recherche Scientifique-FNRS.} 
\and Linda Kleist\thanks{Department of Computer Science, Universität Potsdam, Potsdam, Germany, \texttt{kleist@cs.uni-potsdam.de}.}
\and Michiel Smid\footnotemark[4]
\and Diane Souvaine\thanks{Department of Computer Science, Tufts University, Medford, MA, \texttt{diane.souvaine@tufts.edu}.}
\and Leonidas Theocharous\footnotemark[5]}

\date{}

\newtheorem{theorem}{Theorem}

\newtheorem*{problem*}{Problem}
\newtheorem*{claim*}{Claim}
\newtheorem*{invariant*}{Invariant}
\newtheorem{question}{Question}
\newtheorem{remark}{Remark}

\definecolor{mycolor}{rgb}{0, 0, 0}

\newcommand{\etal}{{et~al.}}
\newcommand{\CH}[1]{\textrm{CH}{(#1)}}
\newcommand{\lin}[2]{\ell{(#1,#2)}}
\newcommand{\ray}[2]{\overrightarrow{#1#2}}

\begin{document}
\maketitle
\begin{abstract}
A {\em covering path} for a finite set $P$ of points in the plane is a polygonal path such that every point of $P$ lies on a segment of  the path. The vertices of the path need not be at points of $P$. A covering path is {\em plane} if its segments do not cross each other. Let $\pi(n)$ be the minimum number such that every set of $n$ points in the plane admits a plane covering path with at most $\pi(n)$ segments.  We prove that $\pi(n)\le \ceil{6n/7}$. This improves the previous best-known upper bound of $\ceil{21n/22}$, due to Biniaz (SoCG 2023). Our proof is constructive and yields a simple $O(n\log n)$-time algorithm for computing a plane covering path.
\end{abstract}

%\todoin{Cyril: The problem with numerical thanks (fixed with footnotes) is that Hugo appears now as "traveling salesperson path problem". Funny! I will try to fix it.}
\setcounter{footnote}{9}

\section{Introduction}

The problems of traversing points in the plane by polygonal paths are fundamental to discrete and computational geometry, and have been studied from both combinatorial and computational points of view. For instance one can refer to traversing a set of points by a non-crossing path \cite{Aichholzer2010,Cerny2007}, or by a path that minimizes the total edge length \cite{Arora1998,Lawler1985,Mitchell002000,Papadimitriou1977},\footnote{Known as the traveling salesperson path problem.} minimizes the longest edge length \cite{BiniazSODA2020,Biniaz2022}, maximizes the shortest edge length \cite{ArkinSODA1997,Arkin1999}, minimizes the number of turns (or bends) \cite{Dumitrescu2014,Stein2001},\footnote{Known as the minimum bend traveling salesperson problem.} or minimizes the total or the largest turning angle \cite{AggarwalSODA1997,Aggarwal1999,Fekete1997}.

\begin{figure}[htb]
	\begin{center}
\includegraphics[width=.2\textwidth]{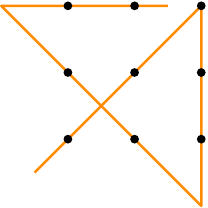}
	\end{center}
    \caption{A non-plane covering path with four segments for the nine-dot puzzle.}
\label{9dot-fig}
\end{figure}

In this paper we focus on traversing a set of points by a non-crossing polygonal path with a small number of segments. This problem has been studied before \cite{BiniazSoCG23,Biniaz2024,Dumitrescu2014,Stein2001}, and is related to the classic nine-dot puzzle that asks to cover the vertices of a $3\!\times\! 3$ grid by a polygonal path with no more than 4 segments \cite{Loyd1914}, as in Figure~\ref{9dot-fig}. 
%It appears in Sam Loyd's  Cyclopedia of Puzzles from 1914 \cite{Loyd1914}.

Let $P$ be a finite set of points in the plane.  A {\em covering path} for $P$ is a polygonal path (i.e., a path with straight-line segments) such that every point of $P$ lies on some segment of the path.
%\todoin{alternative: A {\em covering path} for $P$ is a polygonal curve (or concatenation of segments) containing all points of $P$.}
A vertex of a covering path can be any point in the plane (i.e., not necessarily a point in $P$). The segments of a covering path are also referred to as {\em links} \cite{Arkin2003b} or {\em bends} \cite{Stein2001} in the literature.  A covering path with
the smallest number of links is called a {\em minimum-link} or a {\em minimum-bend} covering path. A covering path is called {\em plane} or {\em non-crossing} if its segments do not cross each other. Figure~\ref{9dot-fig} illustrates a covering path with 4 segments that is not plane.

In addition to their puzzling nature, dating back to 1914~\cite{Loyd1914}, covering paths with a small number of segments are of interest in robotics and heavy machinery, where turning is an expensive operation \cite{Stein2001}.  
Covering paths have also received considerable attention in recent years \cite{Arkin2003b,Biniaz2024,Dumitrescu2014,Keszegh2013}. Mori\'{c} (in CCCG 2010) \cite{Demaine2011} and Welzl (in GWOP 2011) \cite{Demaine2011}, and later Dumitrescu, Gerbner, Keszegh, and
T{\'{o}}th \cite{Dumitrescu2014} raised challenging questions about covering paths, including the following. 

\begin{question}What is the minimum number $\pi(n)$ such that every set of $n$ points in the plane can be covered by a non-crossing path with at most $\pi(n)$ straight-line segments?
\end{question}

It is not hard to see that $n/2 \le \pi(n)\le n{-}1$, because any set of points can be covered by a plane path of $n{-}1$ segments (say by connecting the points from left to right), and each straight-line segment can cover at most two points (if no three points are collinear). Dumitrescu \etal~\cite{Dumitrescu2014} presented the first non-trivial lower and upper bounds:
\[\frac{5n}{9} - O(1)  <\pi(n)  < \pare{ 1-\frac{1}{601\,080\,391} }n.\]
The upper bound has been improved to $\ceil{21n/22}$ by Biniaz~\cite{BiniazSoCG23,Biniaz2024}.
The upper bounds are universal (i.e., any point set admits these bounds), while the lower bounds are existential (i.e., there exist point sets that achieve these bounds).

\subsection{Our Contributions}

We present an algorithm that constructs a plane covering path of at most $\ceil{6n/7}$ segments for any set of $n$ points in the plane, which implies that $\pi(n)\le \ceil{6n/7}$. The algorithm runs in $O(n\log n)$ time. This is optimal, because the problem of computing a plane covering path (even without 
 the minimality constraint) has an $\Omega(n \log n)$ lower bound in the worst case in the algebraic decision tree model of computation---by a  reduction from sorting~\cite{Dumitrescu2014}.

\paragraph*{Simplicity and differences to previous algorithms.} Our algorithm shares similarities with the algorithms of Biniaz~\cite{BiniazSoCG23,Biniaz2024} and Dumitrescu \etal~\cite{Dumitrescu2014} in terms of iteratively scanning the points from left to right. One limitation of the previous approaches is the use of {\em caps} and {\em cups} to save a segment in each iteration. A $k$-{\em cap} is a concave $x$-monotone chain of $k$ points, and a $k$-{\em cup} is a convex $x$-monotone chain of $k$ points. 
Dumitrescu~\etal~construct an 18-cap or 18-cup in each iteration of their  algorithm~\cite{Dumitrescu2014}, while Biniaz uses 5-caps or 5-cups~\cite{BiniazSoCG23,Biniaz2024}. To ensure the existence of such caps or cups, at least 601080391 and 21 points are required, respectively, as implied by the Erd\H{o}s-Szekeres theorem \cite{Erdos1935}. In contrast, our algorithm uses convex chains of points that are not necessarily $x$-monotone (hence neither a cap nor a cup).

Besides using a cap or a cup in each iteration, the previous algorithms~\cite{Biniaz2024,Dumitrescu2014} partition the remaining scanned points (within the iteration) into certain subsets such that each subset is contained in a convex region, and the convex regions of all subsets are disjoint. The algorithms compute a covering path within each region, and then concatenate the paths and the cap or cup.  
%to form a single path. 
The partitioning and the concatenation are nicely crafted so that the resulting path is plane. Our iterations are more straightforward and do not use complex partitioning---we consider three main cases based on the size of the convex hull of 5 points.

\subsection{Related Problems and Results} 

If we allow crossings, one can obtain better upper bounds on the number of segments of covering paths. For example by repeatedly applying the Erd\H{o}s-Szekeres theorem \cite{Erdos1935}, one can obtain a covering path with at most $n/2 + O(n/\log n)$  segments \cite{Demaine2011, Dumitrescu2014}. The optimization problem, which is 
 to compute a minimum-link covering path for a set of points, is shown NP-hard by Arkin \etal~\cite{Arkin2003b}. The best known approximation algorithm, due to Stein and Wagner~\cite{Stein2001}, has a ratio of $O(\log z)$ where $z$ is the
maximum number of collinear points.

A {\em covering tree} for a point set $P$ is a tree with straight-line edges that cover every point of $P$. Covering trees are useful in red-blue separation~\cite{Fulek2013}, and in the construction of rainbow polygons~\cite{Biniaz2024, Flores-Penaloza21}. Let $\tau(n)$ be the minimum number such that every set of $n$ points in the plane can be covered by a plane tree with at most $\tau(n)$ edges. It is known that  $9n/17-O(1)\le \tau(n)\le \ceil{4n/5}$ where the lower and upper bounds come from  \cite{Dumitrescu2014} and \cite{Biniaz2024}, respectively. The exact values of $\pi(n)$ and $\tau(n)$ for the vertices of the square grid have been determined by 
Keszegh \cite{Keszegh2013}.

Two other well-studied related problems to mention are covering points with minimum-link\footnote{For axis-aligned paths this is usually referred to as minimum-turn.} axis-aligned paths \cite{BeregSCG97,Bereg2009,Collins2004,Jiang2015} or with a minimum number of lines \cite{Chen2020,Grantson2006,LangermanESA2002,Langerman2005}.

%\section{NP-hardness}
%\todoin{Ahmad: Apparently the minimum version of the path problem is known to be NP-complete. See open problems number 4, 5, 6 in \cite{Dumitrescu2014}. I believe that during the workshop I mentioned it for trees. For paths I was initially reading \#6, but now that I see \#4, it seems that the harness is known. It looks like \#4 and \#6 contrast each other.}
%\todoin{Ahmad: @Hugo you may add a reduction from polygonization here; a short description would be nice.}
\section{The Algorithm}
In this section we prove the following theorem.

\begin{theorem}
\label{path-thr}
Every set of $n$ points in the plane admits a non-crossing covering path
with at most $\ceil{6n/7}$ segments. Thus, $\pi(n)\leqslant \ceil{6n/7}$. Such a path can be constructed
in $O(n\log n)$ time.
\end{theorem}

Our proof is constructive. We present an algorithm that computes, for any set $P$ of $n$ points in the plane, a non-crossing covering path with at most $\ceil{6n/7}$ segments. 
%%% moved this to where it matters...
%After a suitable rotation we may assume that no two points of $P$ have the same $x$-coordinate.

\paragraph*{Preliminaries.}
We denote the convex hull of a point set $S$ by $\CH{S}$. For two points $p$ and $q$ we denote by $\ell(p,q)$ the line through $p$ and $q$, by $pq$ the line segment with endpoints $p$ and $q$, and by $\ray{p}{q}$ the ray that emanates from $p$ and passes through $q$.
The concatenation of two paths, $\delta_1$ and $\delta_2$, that share an endpoint 
%such that $\delta_1$ ends at the same vertex at which $\delta_2$ starts, we denote their concatenation by
is denoted by 
$\delta_1 \oplus \delta_2$. 

\vspace{10pt}    
Our algorithm is iterative and scans the points from left to right.
Note that via a suitable rotation we may assume that no two points of $P$ have the same $x$-coordinate.
In the first iteration we scan only one point. In each subsequent iteration except for the last, we scan $6$ or $7$ new points and cover them with $5$ or $6$ new segments, respectively. Thus the ratio of the number of new segments to the number of covered points in these {\em intermediate} iterations is at most $6/7$. In the last iteration we scan the (at most $6$) remaining points.

We assume that, among the scanned points in each iteration, no three are collinear. Collinear points are typically easier to handle because one can cover three points by one segment --- such cases can be handled by an argument similar to that of \cite{Biniaz2024}. Let $m$ represent the number of points that have been scanned so far and let $l'$ represent the rightmost scanned point. We maintain the following invariant at the beginning of every iteration after the first. 
\paragraph*{Invariant. } All $m$ points that have been scanned so far are covered by a non-crossing path $\delta$ with at most $\lfloor 6 m/7\rfloor$ segments. The path $\delta$ lies to the left of the vertical line through $l'$. The degree of $l'$ in $\delta$ is one if $m \geqslant 2$ and zero if $m=1$.

\vspace{10pt} 
The invariant holds after the first iteration, trivially.
%as we scan only one point. 
%In what follows, we describe an
We now focus on how to handle each intermediate iteration; this is the main part of our algorithm and proof. The last iteration is described at the end. 
Note that the invariant has the floor-function, whereas Theorem~\ref{path-thr} has the ceiling-function. As we will see, at the beginning of each iteration, the $m$ points scanned so far are covered by at most $\lfloor 6m/7 \rfloor$ segments. After the last iteration, we have $m=n$ and this upper bound becomes $\lceil 6n/7 \rceil$.

In each intermediate iteration the new points are appended to the current path $\delta$ in a suitable way, without altering the structure of $\delta$. 
In our descriptions  we will use $\Delta$ to denote the resulting path. 
%new path $P$ that satisfies the invariant for the next iteration. 

We start by scanning 6 new points. Denote these points by $l$, $a$, $b$, $c$, $r$, $r'$, where $l$ is the leftmost, $r'$ is the rightmost and $r$ is the second rightmost point, as in Figure~\ref{convex3-fig}. Let $S=\{l, a, b, c, r\}$. Thus, $S$ has all newly scanned points except for $r'$. Let $L$ and $R$ be the vertical lines through $l$ and $r$, respectively. We refer to the region between $L$ and $R$ as {\em the slab}. Observe that $l$ and $r$ are both vertices of the boundary of
%the convex hull of $S$, denoted 
$\CH{S}$. We consider three cases depending on the number of vertices on $\CH{S}$. 

%\vspace{20pt}
%\todoin{Note: In order to avoid cross-referencing the figures, we put the description of each case together with its corresponding figures on a separate page.}

%\newpage
\paragraph*{Case 1.}
    $\CH{S}$ has 3 vertices. Without loss of generality, assume that $a$ is the third vertex of $\CH{S}$,  located below $\lin{l}{r}$.
    % The alternate case is symmetric.
    Thus $b$ and $c$ are inside triangle $\bigtriangleup (a,l,r)$. 
    %After a suitable reflection assume that $a$ is below $\lin{l}{r}$.
    The line $\lin{b}{c}$ intersects exactly two edges of $\bigtriangleup (a,l,r)$. We consider three subcases.
    \begin{itemize}
        \item  $\lin{b}{c}$ intersects $\lin{l}{r}$ and $ar$, as shown in Figure~\ref{convex3-fig}(a). The points $l$, $b$, $c$, and $a$ form the vertices of a convex quadrilateral. After a suitable relabeling, we may assume that they appear in this order along the boundary of the quadrilateral. Then the intersection $x$ of $\lin{a}{c}$ and $\lin{l}{b}$ lies in $\bigtriangleup (a,l,r)$. We form the path path $(l', l, x, a, r, r')$ and append it to $\delta$. Thus we cover the six new vertices $(l,b,c,a,r,r')$ with five segments.  The savings come from avoiding using segment $bc$.
\begin{figure}[htb]
	\centering
\setlength{\tabcolsep}{0in}
$\begin{tabular}{ccc}
	\multicolumn{1}{m{.33\columnwidth}}{\centering\vspace{0pt}\includegraphics[width=.28\columnwidth]{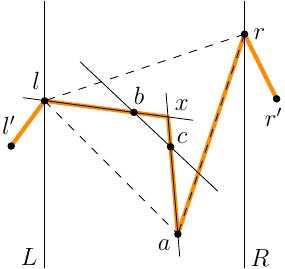}}
		&\multicolumn{1}{m{.33\columnwidth}}{\centering\vspace{0pt}\includegraphics[width=.28\columnwidth]{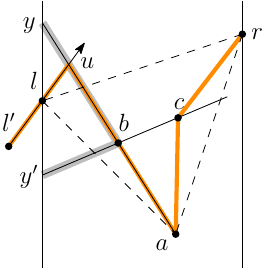}}
        &\multicolumn{1}{m{.33\columnwidth}}{\centering\vspace{0pt}\includegraphics[width=.28\columnwidth]{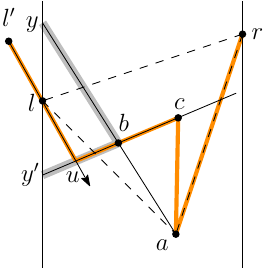}}
		\\
		(a)   &(b) & (c)
\end{tabular}$	
\caption{$\CH{S}$ has three vertices. (a) $\lin{b}{c}$ intersects $lr$ and $ar$. (b) $\lin{b}{c}$ intersects $al$ and $ar$, and $\ray{l'}{l}$ intersects $by$. (c) $\lin{b}{c}$ intersects $al$ and $ar$, and $\ray{l'}{l}$ intersects $by'$.}
\label{convex3-fig}
\end{figure}
 \item $\lin{b}{c}$ intersects $lr$ and $al$. This is symmetric to the previous case. We let $l'$ and $l$ play the role of $r'$ and $r$. Points $a,b,c,r$ form a convex quadrilateral, and we create the path
 $(l', l, a, x, r,r')$.
        \item  $\lin{b}{c}$ intersects $al$ and $ar$. We may assume, without loss of generality, that $b$ is to the left of $c$. Then $\ray{a}{b}$ intersects $L$ or $\ray{a}{c}$ intersects $R$. Because of symmetry, assume that $\ray{a}{b}$ intersects $L$. Their  intersection, $y$, must be  above $l$. Denote by $y'$ the intersection point of $\lin{b}{c}$ with $L$, as in Figure~\ref{convex3-fig}(b). Observe that $y'$ is below $l$.  In this setting $\ray{l'}{l}$ intersects either $by$ or $by'$, at a point $u$. If $\ray{l'}{l}$ intersects $by$,  we set $\Delta=\delta\oplus(l',u,a,c,r,r')$ as in Figure~\ref{convex3-fig}(b), otherwise 
        %if it intersects $by'$, then
        we set $\Delta=\delta\oplus(l',u,c,a,r,r')$ as in Figure~\ref{convex3-fig}(c). Either way we cover  6 vertices with 5
        segments, by avoiding $lb$ on our path.     
    \end{itemize}

\begin{figure}[htb]
	\centering
\setlength{\tabcolsep}{0in}
$\begin{tabular}{ccc}
	\multicolumn{1}{m{.33\columnwidth}}{\centering\vspace{0pt}\includegraphics[width=.28\columnwidth]{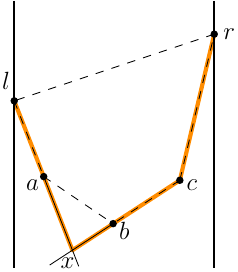}}
		&\multicolumn{1}{m{.33\columnwidth}}{\centering\vspace{0pt}\includegraphics[width=.28\columnwidth]{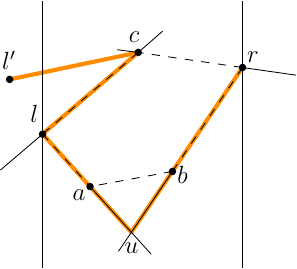}}
        &\multicolumn{1}{m{.33\columnwidth}}{\centering\vspace{0pt}\includegraphics[width=.28\columnwidth]{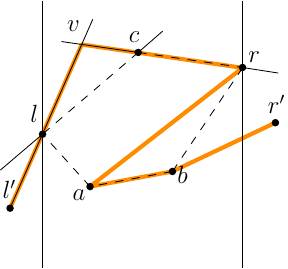}}
		\\
		(a)   &(b) & (c)
\end{tabular}$	
\caption{$\CH{S}$ has five vertices. (a) $a,b,c$ lie on the same side of $lr$. (b) $l'$ is above $\lin{l}{c}$. (c) $l'$ is below $\lin{l}{c}$ and $r'$ is below $\lin{r}{c}$.} 
\label{convex5-fig}
\end{figure}

%\newpage
\paragraph*{Case 2.} $\CH{S}$ has 5 vertices (i.e., $l$, $a$, $b$, $c$, $r$). By  reflection and relabeling, there are two subcases to consider.
    \begin{itemize}
        \item The three vertices $a$, $b$, and $c$ lie on the same side of $lr$ ({\em wlog} below). We may assume that they appear in the order $a,b,c$ from left to right; see Figure~\ref{convex5-fig}(a). Then the intersection point $x$ of $\lin{l}{a}$ and $\lin{b}{c}$ lies in the slab. We set $\Delta=\delta\oplus(l',l,x,c,r,r')$.

        \item Two vertices, say $a$ and $b$, lie on one side of $lr$ ({\em wlog} below), and $c$ lies on the other side. Assume that $a$ is to the left of $b$. 
        The intersection, $u$, of $\lin{l}{a}$ and $\lin{r}{b}$ must be in the slab, by convexity. 
        If $l'$ is above $\lin{l}{c}$ then we set $\Delta=\delta\oplus(l',c,l,u,r,r')$, as shown in Figure~\ref{convex5-fig}(b).  Symmetrically, if $r'$ is above $\lin{r}{c}$, we set $\Delta=\delta\oplus(l',l,u,r,c,r')$.  \\
        Finally if $l'$ is below $\lin{l}{c}$ and $r'$ is below $\lin{r}{c}$, 
        let $v$ be the intersection point of $\lin{l'}{l}$ and $\lin{r}{c}$; see Figure~\ref{convex5-fig}(c). By convexity, $v$ is in the slab, so we can set $\Delta=\delta\oplus(l',v,r,a,b,r')$.
        %Observe that in all cases $x$ lies in the slab.       
    \end{itemize}

%\newpage
\paragraph*{Case 3.} $\CH{S}$ has 4 vertices: $l, r, a, b$. %Let $a$ and $b$ be the other two vertices of $\CH{S}$. 
We consider two subcases.

\begin{figure}[H]
	\centering
\setlength{\tabcolsep}{0in}
$\begin{tabular}{ccc}
	\multicolumn{1}{m{.33\columnwidth}}{\centering\vspace{0pt}\includegraphics[width=.28\columnwidth]{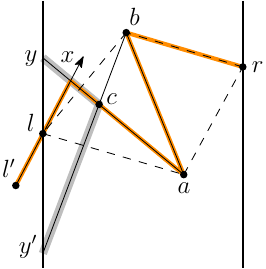}}
		&\multicolumn{1}{m{.33\columnwidth}}{\centering\vspace{0pt}\includegraphics[width=.28\columnwidth]{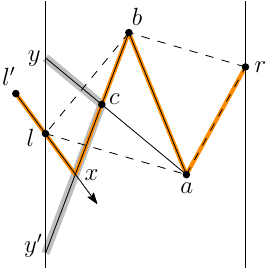}}
        &\multicolumn{1}{m{.33\columnwidth}}{\centering\vspace{0pt}\includegraphics[width=.28\columnwidth]{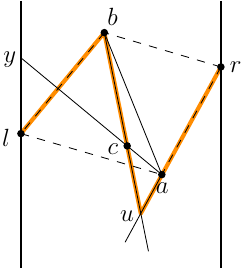}}
		\\
		(a)   &(b) & (c)
\end{tabular}$	
	\caption{$\CH{S}$ has four vertices, the points $a$ and $b$ lie on different sides of $lr$, the point $c$ is in $\bigtriangleup (a,b,l)$, and $\ray{a}{c}$ intersects $L$. (a) $\ray{b}{c}$ intersects $L$, and $\ray{l'}{l}$ intersects $cy$. (b) $\ray{b}{c}$ intersects $L$, and $\ray{l'}{l}$ intersects $cy'$. (c) $\ray{b}{c}$ intersects $R$.}
\label{convex4-1-fig}
\end{figure}

\paragraph*{Case 3a.}
$a$ and $b$ lie on different sides of $lr$. {\em Wlog,} $a$ is below.
%After a suitable reflection, we may assume that $a$ is below and $b$ is above $lr$. 
The point $c$ lies either in triangle $\bigtriangleup(a,b,l)$ or in triangle $\bigtriangleup(a,b,r)$. 
The configurations are symmetric so we show how to handle the former.
%case when $c$ is in $\bigtriangleup(a,b,l)$; the other case is symmetric.
%a reflection of this case and can be handled analogously. 
It is implied that $c$ is to the left of $\ray{a}{b}$, which in turn implies that at least one of $\ray{a}{c}$ or $\ray{b}{c}$ intersects $L$.
{\em Wlog,}  
assume that $\ray{a}{c}$ intersects $L$. The intersection point, $y$, must be  above $l$. There are two subcases to consider:

\begin{itemize}
    \item $\ray{b}{c}$ intersects $L$, at point  $y'$, which must be below $l$. Then either $\ray{l'}{l}$ intersects $cy$  as in  Figure~\ref{convex4-1-fig}(a), or it intersects $cy'$  as in Figure~\ref{convex4-1-fig}(b).
    Either way, let the intersection  point be $x$. 
Respectively,     %intersects $cy$, then 
    we set $\Delta=\delta\oplus(l',x,a,b,r,r')$ 
    %    if it intersects $cy'$, 
    or we set $\Delta=\delta\oplus (l',x,b,a,r,r')$.
    \item $\ray{b}{c}$ intersects $R$, as in Figure~\ref{convex4-1-fig}(c). In this case the intersection point $u$ of $\lin{b}{c}$ and $\lin{r}{a}$ lies in the slab. We set $\Delta=\delta\oplus(l',l,b,u,r,r')$.
\end{itemize}

\begin{figure}[htb]
	\centering
\setlength{\tabcolsep}{0in}
$\begin{tabular}{ccc}
	\multicolumn{1}{m{.33\columnwidth}}{\centering\vspace{0pt}\includegraphics[width=.28\columnwidth]{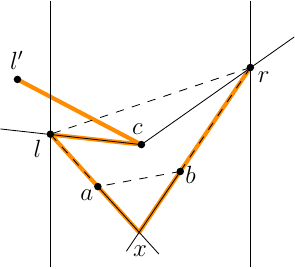}}
		&\multicolumn{1}{m{.33\columnwidth}}{\centering\vspace{0pt}\includegraphics[width=.28\columnwidth]{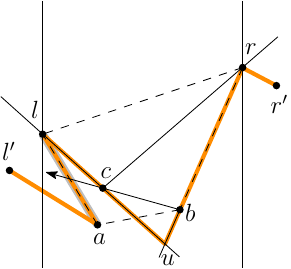}}
        &\multicolumn{1}{m{.33\columnwidth}}{\centering\vspace{0pt}\includegraphics[width=.28\columnwidth]{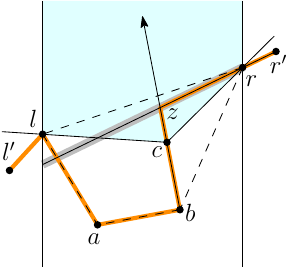}}
		\\
		(a)   &(b) & (c)\\
        \multicolumn{1}{m{.33\columnwidth}}{\centering\vspace{0pt}\includegraphics[width=.32\columnwidth]{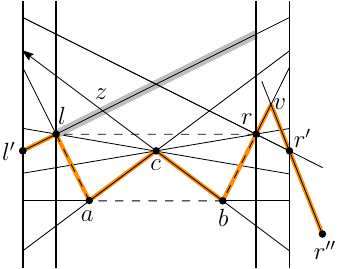}}
		&\multicolumn{1}{m{.33\columnwidth}}{\centering\vspace{0pt}\includegraphics[width=.32\columnwidth]{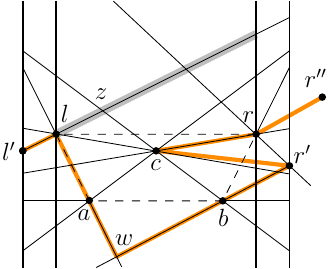}}
        &\multicolumn{1}{m{.33\columnwidth}}{\centering\vspace{0pt}\includegraphics[width=.32\columnwidth]{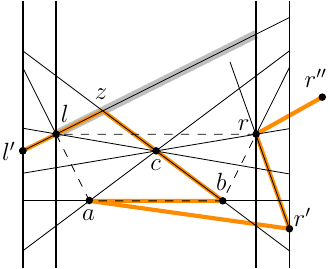}}
		\\
		(d)   &(e) & (f)
\end{tabular}$	
	\caption{$\CH{S}$ has four vertices, $a$ and $b$ lie on the same side of $lr$. (a) $l'$ is above $\lin{l}{c}$. 
    In  (b)-(e) we have $l'$ below $\lin{l}{c}$ and $r'$ below $\lin{r}{c}$. (b) $\ray{b}{c}$ intersects $al$. (c) $\ray{b}{c}$ intersects $\ray{r'}{r}$ within the slab. (d) $r''$ is below $\lin{r}{r'}$. (e) $r''$ is above $\lin{r}{r'}$, and $r'$ is above $\lin{a}{b}$. (f) $r''$ is above $\lin{r}{r'}$, and $r'$ is below $\lin{a}{b}$. } 
\label{convex4-2-fig}
\end{figure}

%\newpage
\paragraph*{Case 3b.}
$a$ and $b$  are on the same side of $lr$, ({\em wlog,} below; and also 
%$a$ and $b$ lie on the same side of $lr$. After suitable reflections, we may assume that $a$ and $b$ lie below $lr$, 
  $a$ is to the left of $b$).
  By convexity, the  intersection point, $x$, of $\lin{l}{a}$ and $\lin{r}{b}$ is in the slab.
If $l'$ lies above $\lin{l}{c}$,  as in Figure~\ref{convex4-2-fig}(a),     we set $\Delta=\delta\oplus(l',c,l,x,r,r')$.
Symmetrically, if $r'$ lies above $\lin{r}{c}$, then we set $\Delta=\delta\oplus(l',l,x,r,c,r')$.
It remains to consider the case when $l'$ is below $\lin{l}{c}$ and $r'$ is below $\lin{r}{c}$. We consider two subcases.

\begin{itemize}
    \item $\ray{b}{c}$ intersects $al$, as in Figure~\ref{convex4-2-fig}(b). In this case, the intersection point $u$ of $\ray{l}{c}$ and $\ray{r}{b}$ is in the slab because $c$ is to the right of $l$, and $b$ is to the left of $r$. Moreover, since $l$ is above  $\ray{b}{c}$, $u$ does not lie on either of the segments $lc$ and $rb$.
    We set $\Delta=\delta\oplus(l',a,l,u,r,r')$. 
    Symmetrically, if $\ray{a}{c}$ intersects $br$ we can set $\Delta=\delta\oplus(l',l,u,r,b,r')$, but this does not seem to help or simplify the next case.
\item $\ray{b}{c}$ does not intersect $al$.  In this case $\ray{b}{c}$ lies in the wedge with counterclockwise boundary ray $\ray{b}{l}$ and clockwise boundary ray $\ray{b}{r}$. 
Because $l'$ is below $\lin{l}{c}$ and $r'$ is below $\lin{r}{c}$, it follows that $\ray{b}{c}$ intersects at least one of the rays $\ray{l'}{l}$ and $\ray{r'}{r}$ within the slab. 
If the intersection is with $\ray{r'}{r}$ then the intersection point $z$ must be inside the shaded region in Figure~\ref{convex4-2-fig}(c), i.e.,
above $lr$ or in the triangle $\bigtriangleup (l,c,r)$. Accordingly,  we set $\Delta=\delta\oplus(l',l,a,b,z,r')$. 
Otherwise, let $z$ be the intersection of $\ray{b}{c}$ with $\ray{l'}{l}$.  This is the case when we need to scan a new point.  \\
Let $r''$ be the next point, after $r'$. If $r''$ is below $\lin{r}{r'}$, then the intersection point $v$ of $\lin{b}{r}$ and $\lin{r''}{r'}$ is in the vertical slab between $r$ and $r'$; we set $\Delta=\delta\oplus(l',l,a,c,b,v,r'')$ as in Figure~\ref{convex4-2-fig}(d). 
Otherwise,  $r''$ is above $\lin{r}{r'}$. Now if $r'$ is above $\lin{a}{b}$, then the intersection point $w$ of $\lin{l}{a}$ and $\lin{r'}{b}$ lies in the slab, below $\lin{a}{b}$; we set $\Delta=\delta\oplus(l',l,w,r',c,r,r'')$ as in Figure~\ref{convex4-2-fig}(e). Because $r'$ is below $\lin{c}{r}$, $cr'$ does not intersect $rr''$. 
If $r'$ is below $\lin{a}{b}$, we set $\Delta=\delta\oplus(l',z,b,a,r',r,r'')$ as in Figure~\ref{convex4-2-fig}(f). In each case we cover seven new points (including $r''$) with six edges.
\end{itemize}

This is the end of an intermediate iteration. In each case, we append to $\delta$ a non-crossing path that lies in the vertical slab between $l$ and the rightmost scanned point. Thus $\Delta$ is non-crossing. Moreover the degree of the rightmost point in $\Delta$ (which is $r'$ or $r''$) is one. In each case we have scanned 6 or 7 new points and covered them by 5 or 6 new segments, respectively. If we have scanned $6$ points, then the total number $M$ of scanned points is $M=m+6$, and the number of segments of $\Delta$ is 
\[|\Delta|=|\delta|+5\leqslant\left\lfloor\frac{6m}{7}\right\rfloor+5= \left\lfloor\frac{6m+35}{7}\right\rfloor\leqslant \left\lfloor\frac{6M}{7}\right\rfloor.\]
If we have scanned $7$ points, then $M=m+7$, and we have 
\[|\Delta|=|\delta|+6\leqslant\left\lfloor\frac{6m}{7}\right\rfloor+6= \left\lfloor\frac{6m+42}{7}\right\rfloor= \left\lfloor\frac{6M}{7}\right\rfloor.\]
Therefore, the invariant holds after every intermediate iteration.\\

In the last iteration of the algorithm we are left with $n'$ points where $1 \leqslant n'\leqslant 6$. We cover these points by an $x$-monotone path with $n'-1$ segments and connect it to the rightmost scanned point (of the path that has been constructed so far) by an additional segment. This gives a non-crossing covering path for all points. The total number of segments in this path is at most 

\[
\floor{\frac{6(n-n')}{7}} + (n' -1)+1 = \floor{\frac{6(n-n')+7n'}{7}} = \floor{\frac{6n+n'}{7}} \leqslant \floor{\frac{6n+6}{7}} \leqslant \ceil{\frac{6n}{7}}.
\]
%\todoin{Cyril: What I understand is that $\pi(n) \le (6n-1)/7$, isn't? ok great, thank you.}
%\todoin{Ahmad: There was a mistake here that I fixed it. Previously we had $n'-1$ which should have been $n'$ with the additional connecting segment. }
Each iteration takes $O(1)$ time. Therefore, after rotating and sorting the points in $O(n\log n)$ time, the rest of the algorithm takes $O(n)$ time. This concludes the proof of Theorem~\ref{path-thr}.

\vspace{8pt}
\begin{remark}
\normalfont
One could scan exactly 7 points in every intermediate iteration, and  obtain the same upper bound. Our choice to  scan 6 points whenever possible serves as a demonstration for the complexity of producing a
non-crossing covering path with at most $5n/6$ segments.
\end{remark}

\vspace{8pt}
\begin{remark}
\label{dif-remark}
\normalfont
In contrast to the algorithms of \cite{Biniaz2024} and \cite{Dumitrescu2014}, our algorithm does not  rely on the existence of caps or cups to reduce the number of segments used. This has been illustrated in Figure~\ref{convex3-fig}(a) when $c$ is to the right of $a$; in Figure~\ref{convex4-2-fig}(c) where $z$ is to the left of $c$; and in Figure~\ref{convex4-2-fig}(f) when $z$ is to the right of $c$. 
\end{remark}

\vspace{8pt}
\begin{remark}
\normalfont
For the $7$ points $l',l,a,b,c,d,r,r'$ in Figure~\ref{convex4-2-fig}(d) we have not found a non-crossing path that starts at $l'$, ends at $r'$, covers all the points, and stays within the slab defined by $l'$ and $r'$. This suggests the necessity of considering $r''$ in our algorithm.  
\end{remark}

\section{Open Problems}

A natural open problem is to narrow the gap between the lower bound $5n/9$ and the upper bound $6n/7$ for $\pi(n)$. 
It is not hard to see that a covering path with a minimum number of segments is a subset of the arrangement of the lines through the $n \choose 2$ pairs of points. Thus one can find a covering path with a minimum number of segments in exponential time. It would be interesting to present a sub-exponential algorithm for computing such a path.

\subsection*{Acknowledgement}
This work was initiated at the 12th  Annual Workshop on Geometry and Graphs, held at the Bellairs Research Institute in Holetown, Barbados, in February 2025. The authors thank the organizers and participants.

\bibliographystyle{plainurl}

\bibliography{Covering-Paths.bib}
\end{document}